\documentclass[11pt,dvips]{article}
\usepackage{graphicx}
\usepackage{amsmath}
\usepackage{amssymb}
\usepackage{latexsym}
\usepackage{color}
\usepackage{cite}
\usepackage{makeidx}
\usepackage{float}
\usepackage{colortbl}
\usepackage{multirow}
\usepackage[dvipdfm,colorlinks]{hyperref}
\newcommand{\bra}{\begin{array}}
\newcommand{\era}{\end{array}}
\newcommand{\beq}{\begin{equation}}
\newcommand{\eeq}{\end{equation}}
\newcommand{\bqr}{\begin{eqnarray}}
\newcommand{\eqr}{\end{eqnarray}}

\def\BC{\bb C}
\def\_\BC{\bbi C}

\def\( {\left(}
   \def\) {\right)}
\def\[ {\left[}
\def\] {\right]}
\def\no2 {{\textstyle{n\over 2}}}

\newcommand{\om}{\omega}

\newcommand{\al}{\alpha}

\newcommand{\lb}{\label}

\begin{document}
\begin{titlepage}
\setcounter{page}{1}
\renewcommand{\thefootnote}{\fnsymbol{footnote}}

\begin{flushright}
\end{flushright}

\vspace{5mm}
\begin{center}

{\Large \bf { Time-dependent Goos-H\"anchen Shifts
in Gapped Graphene}
}

\vspace{5mm}
{\bf Bouchaib Lemaalem}$^{a}$,
 {\bf Miloud Mekkaoui}$^{a}$,
 {\bf Ahmed Jellal}\footnote{a.jellal@ucd.ac.ma}$^{a}$ 
 and {\bf Hocine Bahlouli}$^{b}$

\vspace{5mm}

{$^{a}$\em Laboratory of Theoretical Physics,  
Faculty of Sciences, Choua\"ib Doukkali University},\\
{\em PO Box 20, 24000 El Jadida, Morocco}

{$^b$\em Physics Department,  King Fahd University
of Petroleum $\&$ Minerals,\\
Dhahran 31261, Saudi Arabia}



\vspace{3cm}

\begin{abstract}

We study the Goos-H\"anchen
(GH) shifts for transmitted
Dirac fermions in gapped graphene through a single barrier
structure having a time periodic oscillating
component.
Our
analysis shows that the GH shifts in transmission for central band
$l=0$ and two first sidebands $l=\pm1$ change sign at the Dirac
points $E=V+l\hbar\omega$. In particular the GH shifts in
transmission
exhibit enhanced
peaks at each bound state associated with the single barrier when
the incident angle is less than the critical angle associated with
total reflection.
Klein tunneling, reflected by perfect transmission at normal incidence,
is also preserved in the presence of an oscillating barrier.

\vspace{3cm}



\end{abstract}
\end{center}
\end{titlepage}


\section{ Introduction}

 When a light beam experiences a total reflection at the interface of two 
media having different indices of refraction it shifts along the interface by a certain distance.
Such a lateral displacement between the incident and reflected beams is called 
the Goos-H\"anchen (GH) effect \cite{Goos}. Since the transport carriers in graphene behave like massless particles,
then by analogy with light, GH shift has been investigated in various graphene based nanostructure devices. 
It has also been found that the GH shift can be enhanced by the presence of transmission resonances 
\cite{Chen15,Song16,Chen18} while its control can be achieved through tunability of the applied electrostatic 
potential and induced gap \cite{Chen15,Sharma19}.
It has been found that the GH shift plays an important role in the group velocity of
quasiparticles along interfaces of graphene p-n junctions \cite{Zhao11,Beenakker}.

 Time dependent phenomena and in particular periodic fields and oscillating potential barriers play an important 
role in nanostructure devices. For instances, the application of a periodic oscillating electromagnetic field 
gives rise to additional side band resonant energies at $E+l\hbar \omega$ $(l=0, \pm 1, \cdots)$ in the 
transmission probability \cite{Ahsan,jellal14}. These resonant energies originate from the fact that electrons while interacting with the 
oscillating field will exchange photons of energy $\hbar \omega$, $\omega $ being the frequency of
the oscillating magnetic field. Dayem and Martin \cite{Dayem}  were the first to provide evidence of 
photon assisted tunneling when they subjected a thin superconducting film to a microwave field. Subsequently, 
Tien and Gordon \cite{Tien} used a time modulated scalar potential as a theoretical model to explain these 
experimental observations. Further theoretical studies were
performed later by other research groups, in particular Buttiker
and Landauer investigated the barrier traversal time of particles
interacting with a time-oscillating barrier \cite{Buttiker}. Then
Wagner and other workers \cite{Wagner} gave a detailed treatment
on photon-assisted transport through quantum wells and barriers
with oscillating potentials and studied in depth the transmission
probability as a function of the potential parameters.

In our previous work \cite{jellal14} we have analyzed
the energy spectrum together with the corresponding transmission
in graphene 
through a square potential barrier driven by a
periodic potential. In the present work
we consider the same system but with a gap and study
%
%
%
 and study the transport of Dirac fermions. 
The barrier height oscillates sinusoidally
around an average value $V_j$ with oscillation amplitude $U_j$ and
frequency $\omega$. Thus we will investigate the negative and
positive GH shifts in transmission for the central band and
sidebands of Dirac fermions through a time-oscillating potential
in monolayer graphene, based on the tunable transmission gap
\cite{ChenX, Chen15}. The GH shifts for the central band and first
sidebands discussed here are related to the transmission
resonances, which are quite different from the GH shift for total
reflection at a single graphene interface. Using the derived energy spectrum we compute the GH 
shifts for the central and sidebands and study their variations in terms of the system 
physical parameters and phase shifts.
To give a better understanding of our results, we perform a numerical study based on
various choices of the physical parameters. Among the obtained
results we show that GH shifts in transmission can be controlled
by a square potential barrier harmonically oscillating in time.
{It is worth mentioning that few recent references have considered
the GH shifts due to irradiation of graphene sheet with a time-dependent oscillating magnetic field \cite{Huaman}. Others investigated the GH shifts in a strained graphene  sheet where the mechanical strain is described by a gauge vector potential giving rise to a pseudo-magnetic field that affected differently the valley \cite{Zhen}. 
}

 The work is organized as follows. In section $2$, we
 use the solutions of the energy spectrum associated with our system
 together with transmission probabilities to
 determine the GH shifts.
  We numerically analyze and discuss the GH shifts in
transmission within the central band and first sidebands by considering
suitable choices of the physical parameters in section 3. Conclusions are given
in last section.

\section{ Goos-H\"anshen shifts}

Opening and controlling a band gap in graphene is one
of the most important issues that need to be resolved with 
certitude to ensure the usage of graphene in telecommunication 
or information technology. Several methods have been advanced to create such 
band gap in graphene such as deposition of graphene on a well
selected substrate having a similar honeycomb structure \cite{Guinea},
application of strain \cite{Choi} or creation of nanoribbons by 
physically cutting the graphene sheet  giving rise to 
an effective mass \cite{Son}. However the existence of a substrate 
induced energy gap in graphene due to its interaction with substrate  SiC or BN is still debatable \cite{Rotenberg}. Setting aside the issue of existence 
and realization of an energy gap in graphene while preserving its honeycomb lattice symmetry, we consider in our model a gapped graphene
subject to 
a square potential barrier of
width $d$ and oscillating sinusoidally around its average height  $V_{j}$ with amplitude $U_{j}$ and frequency
$\omega$. Fermions with energy $E$ are incident from one side of
the barrier at incidence angle $\phi_{0}$ with respect to the $x$-axis
and leave the barrier with energy $E+ m\hbar \omega$ $(m=0, \pm
1, \pm 2, \cdots)$ making angles $\pi-\phi_{m}$ after reflection
and $\phi_{m}$ after transmission. The corresponding Hamiltonian
is
\begin{equation}\lb{ham1}
H_{j}=-i\hbar v_{F}\ \vec \sigma\cdot\vec \nabla 
+\Delta\sigma_{z}+ \left[V_{j}(x)+ U_{j}\cos(\omega t)\right] {\mathbb I}_{2}
\end{equation}
where $\upsilon_{F}$ is  the Fermi velocity, $ \sigma =(\sigma_{x}, \sigma_{y})$ are  the Pauli matrices, ${\mathbb I}_{2}$ is
the $2 \times 2$ unit matrix,  $V$ is the static
square potential barrier and  $U_{j}$ is the amplitude of the oscillating
potential,  {both of which are constant} for
$0\leq x\leq d$, with $d$ positive, and  zero elsewhere (Figure \ref{fig0}) 
\begin{equation}
V_{j}(x)=
\left\{%
\begin{array}{ll}
    V, & \qquad\hbox{$0\leq x\leq d$} \\
    0, & \qquad \hbox{otherwise} \\
\end{array}%
\right.,\qquad U_{j}=
\left\{%
\begin{array}{ll}
    u_{1}, & \qquad \hbox{$0\leq x\leq d$} \\
    0, & \qquad \hbox{otherwise} \\
\end{array}%
\right.
\end{equation}
The subscript $j = {\sf 0}, {\sf 1}, {\sf 2}$ denotes each
scattering region from left to right
as shown in Figure \ref{fig0}. The parameter $\Delta = m v_{F}^2$ is the
energy gap owing to the sublattice symmetry breaking.

\begin{figure}[!ht] \centering
\includegraphics[width=10cm, height=4cm]{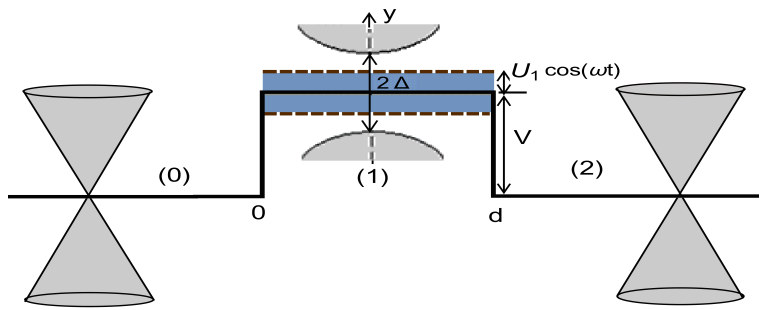}
 \caption{\sf{Schematic of  the potential profile showing a time oscillating potential in a 
 gapped graphene region subject to an electrostatic potential.}}\lb{fig0}
\end{figure}

After rescaling $\epsilon =
E/v_{F}$, $v_{j}= V_{j}/v_{F}$, $\mu= {\Delta}/v_{F}$,
$\varpi=\omega/v_{F}$, $u_{j}= U_{j}/v_{F}$, $\al_1=\frac{u_1}{\varpi}$
and taking $\hbar=1$,
we solve the Dirac equation
in the three regions. {In the incident region $j={\sf 0}$} $(x<0)$:
\begin{eqnarray}
\psi_{\sf
0}(x,y,t)=e^{ik_{y}y}\sum^{+\infty}_{m,l=-\infty}\left[\delta_{l,0}\left(
\begin{array}{c}
1 \\
 z_{l}\end{array}\right)e^{ik_{l}x}+r_{l}\left(
\begin{array}{c}
1 \\
 -\frac{1}{z_{l}}\end{array}\right)e^{-ik_{l} x
 }\right]\delta_{m,l}e^{-iv_{F}(\epsilon+m\varpi)t}
\end{eqnarray}
where $z_{l}=s_{l}\frac{k_{l} +ik_{y}}{\sqrt{k^{2}_{l} +k_{y}^{2}}}$
and  $s_{l}=\mbox{sgn}(\epsilon+l \varpi)$.
{In the transmitted region $j={\sf 2}$} $(x>d)$:
\begin{equation}
\psi_{\sf
2}(x,y,t)=e^{ik_{y}y}\sum^{+\infty}_{m,l=-\infty}\left[t_{l}\left(
\begin{array}{c}
1 \\
 z_{l}\end{array}\right)e^{ik_{l} x
 }+b_{l}\left(
\begin{array}{c}
1 \\
 -\frac{1}{z_{l}}\end{array}\right)e^{-ik_{l} x}\right]\delta_{m,l}e^{-iv_{F}(\epsilon+m\varpi)t}
\end{equation}
where $\{b_{l}\}$ is the null vector. {In the scattering region 
$j={\sf 1}$} $(0<x<d)$:
\begin{eqnarray}
 \psi_{\sf 1}(x,y,t)
=e^{ik_{y}y}\sum^{m,l=+\infty}_{m,l=-\infty}\left[a_{l}^{j}\left(
\begin{array}{c}
c_+ \\
 c_- z_{l}^{'}\end{array}\right)e^{ik^{'}_{l}x}+b_{l}^{j}\left(
\begin{array}{c}
c_+ \\
 -\frac{c_-}{z_{l}^{'}}\end{array}\right)e^{-ik^{'}_{l} x}\right]
J_{m-l}(\al_1)e^{-iv_{F}(\epsilon+m\varpi)t}
\end{eqnarray}
where
$ c_{\pm}=\sqrt{1 \pm \frac{s_{l}^{'}\mu}{\sqrt{\left(k_{l}^{'}\right)^2+{k_{y}}^{2}}}},
z_{l}^{'}=s_{l}^{'}\frac{k_{l}^{'}+ik_{y}}{\sqrt{\left(k_{l}^{'}\right)^2+k_{y}^{2}}}
$
and the associated energy
$
 \left(\epsilon-v+l\varpi\right)^{2}-\mu^{2}=
 s_l'\sqrt{\left(k_{l}^{'}\right)^2+{k_{y}}^{2}}
$,
with
$
s_{l}^{'}=\mbox{sgn}(\epsilon+l \varpi-v)
$ and
$J_{m}(\al_1)$
is the Bessel function of the first kind 
{such that 
$J_{m-l}
(\al_1)=\delta_{m,l}$.}


 The transmission and reflection probabilities
 can be obtained using the continuity of
the spinor wavefunctions at junction interfaces $(x=0, x=d)$, namely
  $  \psi_{\sf 0}(0,y,t)=\psi_{\sf 1}(0,y,t)$ and
   $\psi_{\sf 1}(d,y,t)=\psi_{\sf 2}(d,y,t)$.
These {boundary conditions can be represented 
in matrix form as}
\begin{eqnarray}\lb{Mat1}
\left(%
\begin{array}{c}
  \Xi_{0} \\
  \Xi_{0}^{'} \\
\end{array}%
\right)=\left(%
\begin{array}{cc}
 { \mathbb M_{11}} &{\mathbb M_{12}} \\
 {\mathbb M_{21}} &{ \mathbb M_{22}} \\
\end{array}%
\right)\left(%
\begin{array}{c}
  \Xi_{2} \\
  \Xi_{2}^{'}\\
\end{array}%
\right)={\mathbb M}\left(%
\begin{array}{c}
  \Xi_{2} \\
 \Xi_{2}^{'} \\
\end{array}%
\right)
\end{eqnarray}
and the total transfer matrix ${\mathbb M}={\mathbb
M(0,1)}\cdot{\mathbb M(1,2)}$ with  
\begin{align}
{\mathbb M(0,1)}=\left(%
\begin{array}{cc}
  {\mathbb I}& {\mathbb I} \\
{\mathbb N^{+}} &{\mathbb N^{-}} \\
\end{array}%
\right)^{-1}
\left(%
\begin{array}{cc}
  {\mathbb C_{1}} & {\mathbb C_{1}} \\
 {\mathbb G_{1}^{+}} & {\mathbb G_{1}^{-}} \\
\end{array}%
\right), \ \
{\mathbb M(1,2)}=\left(%
\begin{array}{cc}
  {\mathbb C_{2}^{+}} & {\mathbb C_{2}^{-}} \\
  {\mathbb G_{2}^{+}} & {\mathbb G_{2}^{-}} \\
\end{array}%
\right)^{-1}
\left(%
\begin{array}{cc}
  {\mathbb I}& {\mathbb I} \\
{\mathbb N^{+}} &{\mathbb N^{-}} \\
\end{array}%
\right)\left(%
\begin{array}{cc}
  {\mathbb K^{+}}&{\mathbb O}  \\
{\mathbb O} &{\mathbb K^{-}} \\
\end{array}%
\right)
\end{align}
where we have set the parameters
$
\left({\mathbb N^{\pm}}\right)_{m,l}=\pm\left(z_{m}\right)^{\pm
1}\delta_{m,l},
\left({\mathbb
C_{1}^{\pm}}\right)_{m,l}=\alpha_{l}J_{m-l}\left(\frac{u_{1}}{\varpi}\right),
\left({\mathbb G_{1}^{\pm}}\right)_{m,l}=\pm\beta_{l}
(z_{l}^{'})^{\pm 1}J_{m-l}\left(\frac{u_{1}}{\varpi}\right),
\left({\mathbb C_{2}^{\pm}}\right)_{m,l}=\alpha_{l}e^{\pm
ik_{l}^{'}d}J_{m-l}\left(\frac{u_{1}}{\varpi}\right), 
\left({\mathbb G_{2}^{\pm}}\right)_{m,l}=\pm\beta_{l}
(z_{l}^{'})^{\pm 1}e^{\pm
ik_{l}^{'}d}J_{m-l}\left(\frac{u_{1}}{\varpi}\right),
\left({\mathbb  K^{\pm}}\right)_{m,l}=\pm e^{\pm
idk_{l}}\delta_{m,l}$
and
the null matrix is denoted by ${\mathbb O}$,   ${\mathbb I}$ is
the unit matrix.

To proceed further,
we assume an electron propagating from left to
right with the quasienergy $\epsilon=\frac{E}{v_{F}}$. Then, $\tau=(1,2)$,
$\Xi_{0}=\{\delta_{0,l}\}$ and $\Xi_{2}^{'}=\{a_{m}\}$ is the null
vector, whereas $\Xi_{2}=\{t_{l}\}$ and $\Xi_{0}^{'}=\{r_{l}\}$
are the vectors for transmitted  and  reflected waves,
respectively, $\Xi_{2}=\left({ \mathbb M_{11}}\right)^{-1}\cdot \Xi_{0}$.
The minimum number $N$ of sidebands that need to be considered is
determined by the strength of the perturbing potential oscillation,
$N>\frac{v_{1}}{\varpi}$. The infinite series for $T$ can be {then}
truncated to consider {only} a finite number of terms starting from $-N$
up to $N$. Furthermore, analytical results are obtained if we
consider small values of $\alpha_1=\frac{u_{1}}{\varpi}$ and
include only the first two sidebands at energies $\epsilon\pm
\varpi$ along with the central band at energy $\epsilon$
\begin{equation}\lb{tttt}
t_{-N+k}=\mathbb{M'}\left[k+1, N+1\right]
\end{equation}
where $k=0, 1, 2,\cdots, 2N$ and ${ \mathbb M^{'}}$ is a matrix
element of ${ \mathbb M_{11}}^{-1}$. Using the reflected
$J^{\sf {re}}$ and transmitted $J^{\sf {tr}}$ currents,
 the reflection $R_{l}$ and transmission $T_{l}$ probabilities
 for a given  mode $l$ can be expressed as
\begin{equation}
  T_{l}= \frac{k_{l}}{k_{0}}
  |t_{l}|^{2},\qquad
  R_{l}= \frac{k_{l}}{k_{0}}
  |r_{l}|^{2}
\end{equation}
such that $T_{l}$ is the probability amplitude describing the
scattering of an electron with incident quasienergy $E$ in the
region ${\sf 0}$ into the sideband with quasienergy $E+l\hbar\omega$
in the region ${\sf 1}$. Thus, the rank of the transfer matrix
${\mathbb M}$ increases with the amplitude of the
time-oscillating potential. The amplitudes $t_l$ and $r_l$
can be written in complex notation as
$
t_l = \rho_{l}^{t}\ e^{i \varphi_{l}^{t}},
r_l = \rho_{l}^{r}\ e^{i \varphi_{l}^{r}}
$
such that the corresponding  phase shifts and {moduli are
defined by}
$
\varphi_{l}^{t} = \arctan\left(i\frac{t_{l}^{\ast}-t_{l}}{t_{l}+t_{l}^{\ast}}\right),
  \varphi_{l}^{r} =
 \arctan\left(i\frac{r_{l}^{\ast}-r_{l}}{r_{l}+r_{l}^{\ast}}\right),
     \rho_{l}^{t} =\sqrt{\text{Re}^2 [t_{l}]+\text{Im}^2 [t_{l}]},
     \rho_{l}^{r} =\sqrt{\text{Re}^2 [r_{l}])+\text{Im}^2 [r_{l}]}
 $,
 which
can be used to obtain the probabilities
\begin{equation}
  T_{l}=\frac{k_{l}}{k_{0}}\left(\text{Re}^2 [t_{l}]+\text{Im}^2 [t_{l}]\right),\qquad  R_{l}=\frac{k_{l}}{k_{0}}\left(\text{Re}^2 [r_{l}])+\text{Im}^2 [r_{l}]\right).
\end{equation}
In the forthcoming analysis due to numerical difficulties, we are able to truncate 
 \eqref{tttt} retaining only the terms corresponding to
the central and first two sidebands, namely $l =0, \pm 1$. We can
proceed as before to derive transmission amplitudes
$
  t_{-1}=\mathbb{M^{'}}[1,2],  t_{0}=\mathbb{M^{'}}[2,2], 
t_{1}=\mathbb{M^{'}}[3,2].
$
Such an approximation can be validated at low energies where two 
and higher-photon processes are less probable than the single photon processes.

The Goos-H\"anchen shifts for Dirac fermions in gapped graphene under the applied potential can be analyzed by
considering the incident, reflected and transmitted beams around
some transverse wave vector $k_y = k_{y_0}$ and the angle of
incidence $\phi_{l}(k_{y_{0}})$ lies in the interval $[0, \frac{\pi}{2}]$.
These beams can be written in terms of the obtained solutions of the
 energy spectrum. Indeed, for the incident and reflected waves, {we have}
\begin{eqnarray}
  && \Psi_{\sf in}(x,y) = \int_{-\infty}^{+\infty}dk_y\ f(k_y-k_{y_0})\ e^{i(k_{0}(k_y)x+k_yy)}\left(
            \begin{array}{c}
              {1} \\
              {e^{i\phi_{0}(k_{y})}}
            \end{array}
          \right)\label{eq 79}\\
&&
\Psi_{\sf re}(x,y) =\int_{-\infty}^{+\infty}dk_y\ r_{l}\ 
f(k_y-k_{y_0})\ e^{i(-k_{l}(k_y)x+k_yy)}\left(
            \begin{array}{c}
              {1} \\
              {-e^{-i\phi_{l}(k_{y})}} \\
            \end{array}
          \right)\label{refl}
\end{eqnarray}
where 
$\phi_l=\tan^{-1}\frac{k_y}{k_l}$,
 $\phi_0$ is the incident angle 
and $f(k_y - k_{y0})$  the angular spectral
distribution. We can approximate the $k_y$-dependent terms by a
Taylor expansion around $k_{y_0}$, retaining only the first order term
to end up with
\begin{eqnarray}
\phi_{l}(k_{y})\approx
\phi_{l}(k_{y_{0}})+\frac{\partial\phi_{l}}{\partial
k_{y}}\Big|_{k_{y_{0}}}(k_{y}-k_{y_{0}}), \qquad
 k_{l}(k_{y})\approx k_{l}(k_{y_{0}})+\frac{\partial
k_{l}}{\partial k_{y}}\Big|_{k_{y_{0}}}(k_{y}-k_{y_{0}}).
\end{eqnarray}
As for the transmitted beam, we have 
\begin{eqnarray}
\Psi_{\sf tr}(x,y) &=& \int_{-\infty}^{+\infty}dk_y\ t_{l}\ 
f(k_y-k_{y_0})\ e^{i(k_{l}(k_y)x+k_yy)}\left(
            \begin{array}{c}
              {1} \\
              {e^{i\phi_{l}(k_{y})}} \\
            \end{array}
          \right)\label{trans}.
\end{eqnarray}

 The stationary-phase approximation indicates that the GH shifts are equal to the negative gradient
of transmission phase with respect to $k_y$. To calculate the GH
shifts of the transmitted beam through our system we adopt the definition from
\cite{Chen1, mMekkaoui}, accordingly the stationary phase method \cite{Bohm} gives
\begin{equation}
        S_{l}^{t}=- \frac{\partial \varphi_{l}^{t}}{\partial
        k_{y}}\Big|_{k_{y0}}, \qquad S_{l}^{r}=- \frac{\partial \varphi_{l}^{r}}{\partial
        k_{y}}\Big|_{k_{y0}}.\label{eq 51}
 \end{equation}
Assuming a finite beam width with  Gaussian shape,
$f(k_y-k_{y_0})=w_y\exp[-w_{y}^2(k_y-k_{y_0})^2]$ around $k_{y0}$,
where $w_{y}=w\sec\phi_l$,  with  half beam width $w$ at the
waist, we can evaluate the Gaussian integral to obtain the spatial
profile of the incident beam, by expanding $\phi_l$ and $k_{l}$ to
first order around $k_{y0}$ when satisfying the {required} condition
$
\delta\phi_l=\lambda_{F}/(\pi w)\ll 1
$
with  the Fermi wavelength $\lambda_{F}$. Comparison of the
incident and transmitted beams suggests that the displacements
$\sigma_{\pm}$ of up and down spinor components are both equal to
$\partial \varphi_{l}^{t}/\partial k_{y0}$ and the average
displacement is {given by}
\begin{equation}
S_{l}^{t}=\frac{1}{2}(\sigma^{+}+\sigma^{-})=- \frac{\partial
\varphi_{l}^{t}}{\partial
        k_{y}}\Big|_{k_{y0}}.
\end{equation}
%
 Next we will numerically analyze and discuss the GH shifts for the central band $S_{0}^{t}$ and first sidebands $S_{\pm 1}^{t}$ for
Dirac fermions in gapped graphene scattered by a square barrier with height that
oscillates sinusoidally. This will be done by {selecting adequately}
 the
physical parameters characterizing our system.

\section{Discussion of numerical results}

To allow for a suitable interpretation of our main results, we
compute numerically the GH shifts in transmission for the central band
and first sidebands under various conditions. First we plot the GH
shifts  in transmission $S_{l}^{t}$  as a
function of the  potential $v$ of the oscillating barrier
in the gapless graphene {region where} $\mu=0$, the energy $\epsilon=10$ and
the frequency $\varpi=1$, see Figure \ref{figm1}. It is clear  
that the GH shifts change sign at the Dirac points,
namely $v =\epsilon+l\varpi$ with $(l=-1, 0, 1)$.
We observe that $S_{l}^{t}$ exhibit negative as well as
positive values and strongly {depend} on the location of Dirac
points.
  In Figure \ref{figm1}(a), we observe,
  {when}
   $d=2.5$, $\alpha=0.2$ and $k_{y}=1$,
  that the GH shifts
  in transmission for the central band  and the two first sidebands
  $S_{0}^{t}$ (blue line), $S_{-1}^{t}$ (green line) and $S_{1}^{t}$
  (red line) change sign at the Dirac points $\epsilon$,
  $\epsilon-\varpi$ and $\epsilon+\varpi$, respectively.
  Figure~\ref{figm1}(b) shows
for different values of $\alpha_1=\{0.2, 0.6, 0.9\}$,
that the GH shifts for  central band $S_{0}^{t}$ in the oscillating barrier
decreases if $\alpha_1$ increases. This tells us
that by adjusting
the value of $\alpha_1$ we can decrease the value of
$S_{0}^{t}$. 
In Figure \ref{figm1}(c),
we have chosen the parameters
$\alpha_1=0.2$, $k_{y}=1$ for three different values of the
distance $d=0.5$ (red line), $d=1.5$ (green line), $d=2.5$ (blue
line) to show $S_{0}^{t}$ behaviors.  We  observe that
$S_{0}^{t}$ increase if $d$ increases
and change its sign
at the Dirac points $v=\epsilon$.
This change in sign of the GH shifts shows
clearly that they  strongly depend on the strength of the barrier heights.
Note that, the Dirac points represent the zero modes for Dirac operator
\cite{Sharma19} and lead to the emergence of new Dirac points, which
have been discussed in different works \cite{Wang,Chen1}. Such
point separates the two regions of positive and negative
refraction. In the cases of $v<\epsilon$ and $v>\epsilon$, $S_{0}^{t}$ is, respectively, in the forward and
backward directions due to the fact that the signs of group
velocity are opposite.
Figure \ref{figm1}(d) presents the numerical results of
the GH shifts in transmission $S_{1}^{t}$ for first band $l=1$ as a function
of the  potential energy $v$ for specific values of the barrier
width $d=1.5$, $\alpha_1=0.4$ and three different values of the
wavevector $k_{y}=0$ (red line), $k_{y}=2$ (green line), $k_{y}=4$
(blue line). We observe that
$S_{1}^{t}$ decrease if $d$ decreases and then vanish for
$k_y=0$, that is to say for normal incidence there is no shift.

\begin{figure}[!ht] \centering
\includegraphics[width=6cm, height=3.3cm]{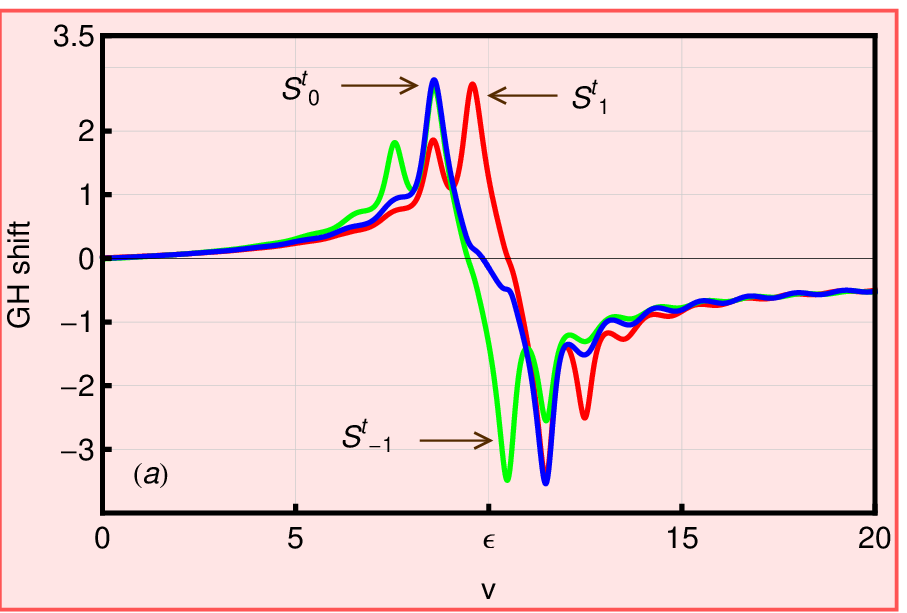}\ \ \ \
\includegraphics[width=6cm, height=3.3cm]{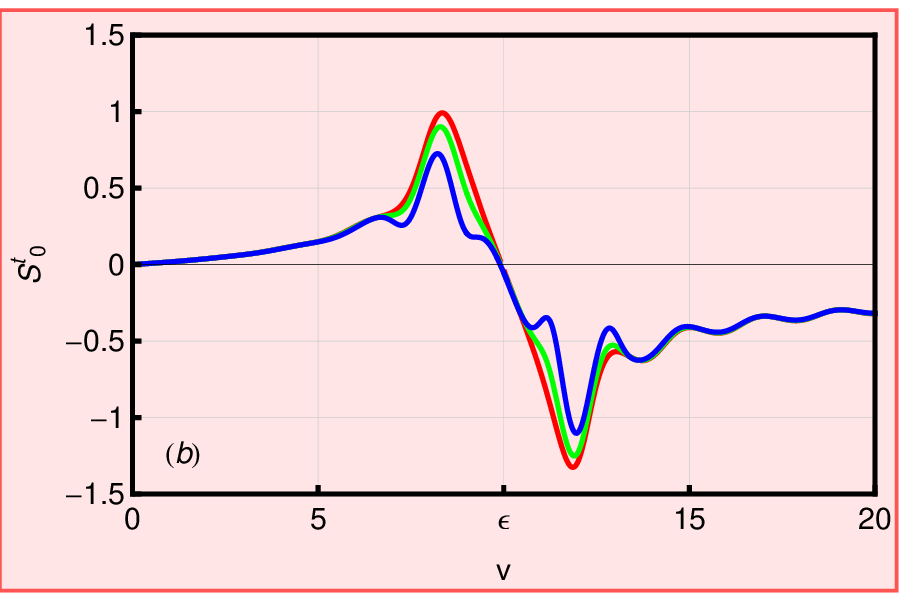}\\
\includegraphics[width=6cm, height=3.3cm]{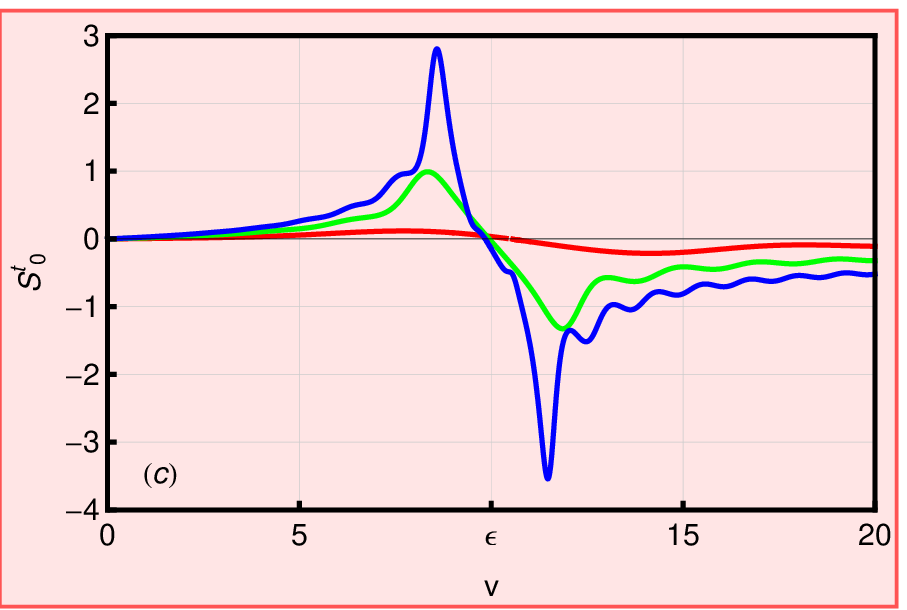}\ \ \ \
\includegraphics[width=6cm, height=3.3cm]{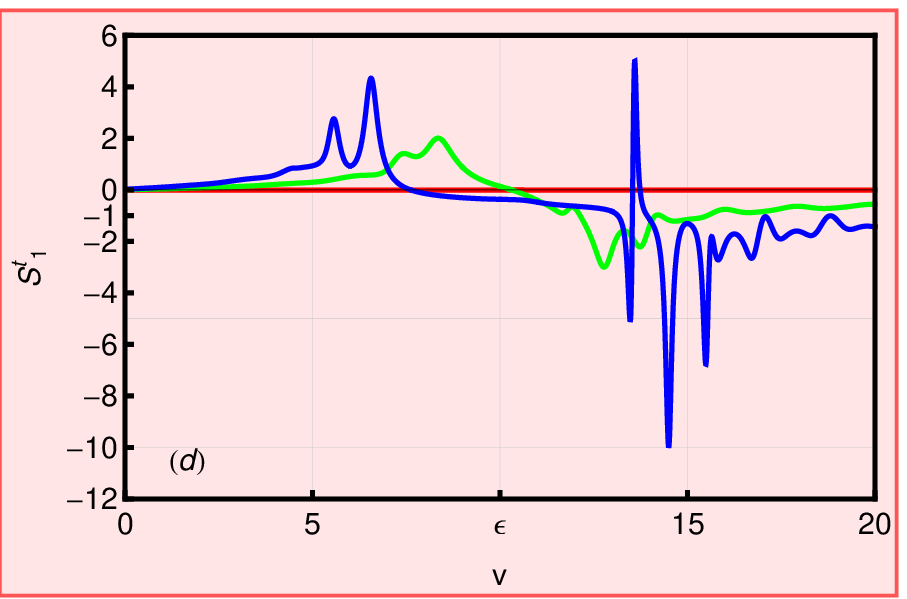}
 \caption{\sf{(color online) GH shifts in transmission $S_{l}^{t}$ versus potential $v$ for the oscillating
 barrier with $\mu=0$, $\epsilon=10$, $\varpi=1$.  (a): $l={-1, 0, 1}$, $k_{y}=1$,  $d=2.5$, $\alpha_1=0.2$. (b): $l=0$, $k_{y}=1$,  $d=1.5$ $\alpha_1=0.2$ (red
line), $\alpha=0.6$ (green line), $\alpha=0.9$ (blue line). (c):
$\alpha_1=0.2$, $l=0$, $k_{y}=1$, $d=0.5$ (red line), $d=1.5$ (green
line), $d=2.5$ (blue line). (d): $\alpha_1=0.4$,  $l=1$, $d=1.5$
$k_{y}=0$ (red line), $k_{y}=2$ (green line), $k_{y}=4$ (blue
line).}}\lb{figm1}
\end{figure}

In Figure \ref{figm2} we present the numerical results of the
transmission probabilities and GH shifts in transmission as a
function of barrier width $d$ with $\epsilon=10$, $v=15$, $\mu=0$,
$\varpi=1$, $k_{y}={0, 2}$. The transmission probability $T_{0s}$
and the GH shifts $S_{0s}^{t}$ for the static barrier are shown, which  correspond to the case {$\alpha_1=\frac{u_1}{\om}=0$}. While for the oscillating
barrier with $\alpha_1=0.6$, we show 
the GH shifts in
transmission $S_{l}^{t}$  and transmission probabilities $T_{l}$
for central band $l=0$ and first sidebands $l=\pm 1$ as a function
of barrier width. The magenta, blue, green and red lines
correspond to ($T_{0s}$, $S_{0s}^{t}$), ($T_{0}$, $S_{0}^{t}$),
($T_{-1}$, $S_{-1}^{t}$) and ($T_{1}$, $S_{1}^{t}$),
respectively.
  Figures \ref{figm2}(a,c) present the
transmission probabilities and GH shifts in transmission as
function of barrier width $d$ for the wavevector $k_{y}=2$ 
where the transmission
probabilities for the central band and sidebands in the
oscillating barrier show sinusoidal behaviors. We observe that the number of
peaks in the GH shifts for the first sidebands $S_{-1}^{t}$ and
$S_{1}^{t}$ correspond to the zero transmission probabilities for the
first sidebands $T_{-1}$ and  $T_{1}$, respectively.
  Figures \ref{figm2}(b,d) show  the transmission probabilities and
 GH shifts in transmission for normal incidence $k_y=0$. Obviously,  $T_{0s}$ is unity for larger barrier width and the GH shifts
 $S_{0s}^{t}$ are zero. For the oscillating barrier,
 $T_{0}$ varies initially from unity and oscillates
periodically for larger barrier width. However, the transmission
probabilities for  the other first two sidebands $T_{\pm 1}$
{starts} initially from zero  then oscillates periodically. This occurs
due to the larger tunneling time available for the electron to interact with the
oscillating potential as it traverses the barrier. In addition, we
find that for normal incidence in the oscillating barrier
$T_{+1}=T_{-1}$ and the GH shifts $S_{\pm 1}^{t}=0$. Moreover, the
total transmission probability through the central band as well as
the sidebands is unity. These results imply that perfect
transmission at normal incidence is independent of the barrier
width, which is yet another manifestation of Klein tunneling.

\begin{figure}[!ht] \centering
\includegraphics[width=6cm, height=3.3cm]{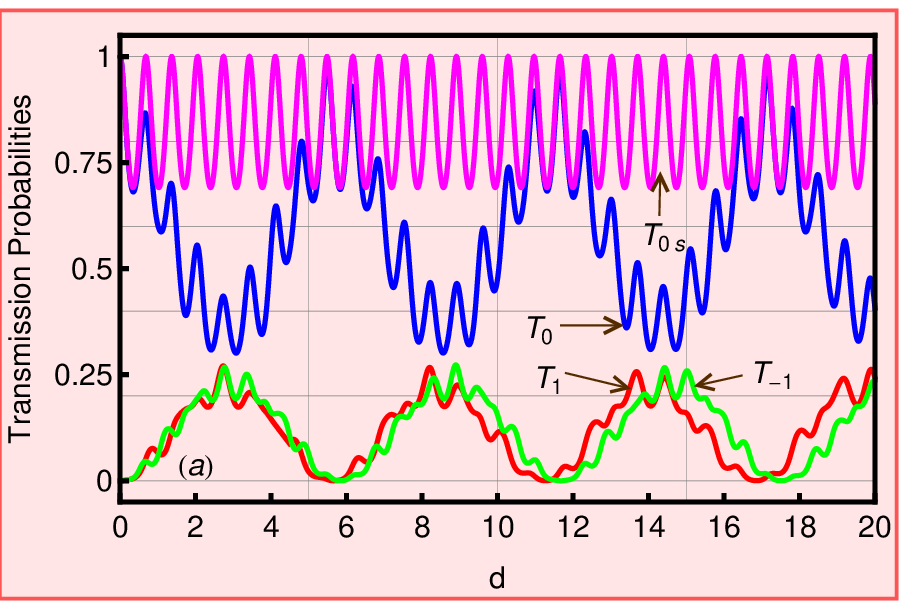}\ \ \ \
\includegraphics[width=6cm, height=3.3cm]{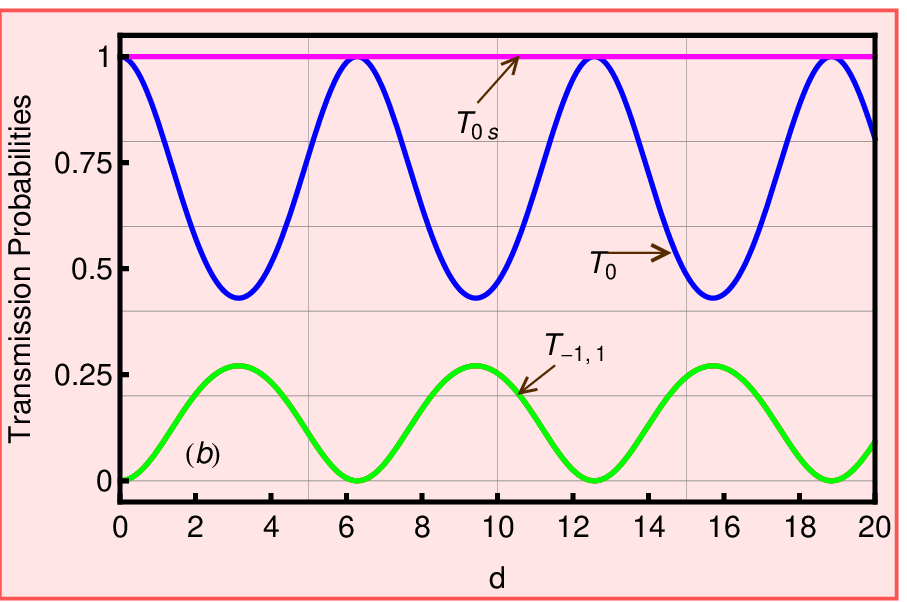}\\
\includegraphics[width=6cm, height=3.3cm]{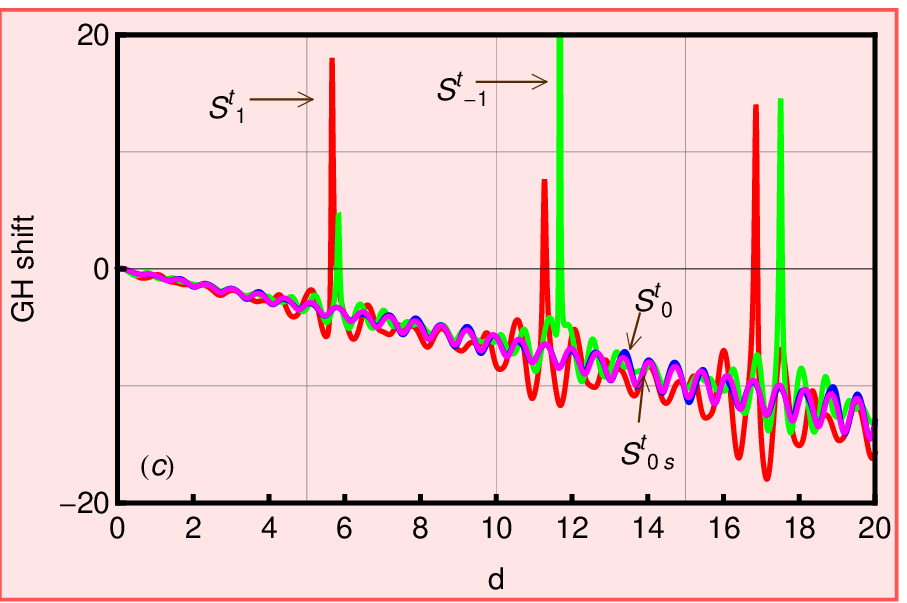}\ \ \ \
\includegraphics[width=6cm, height=3.3cm]{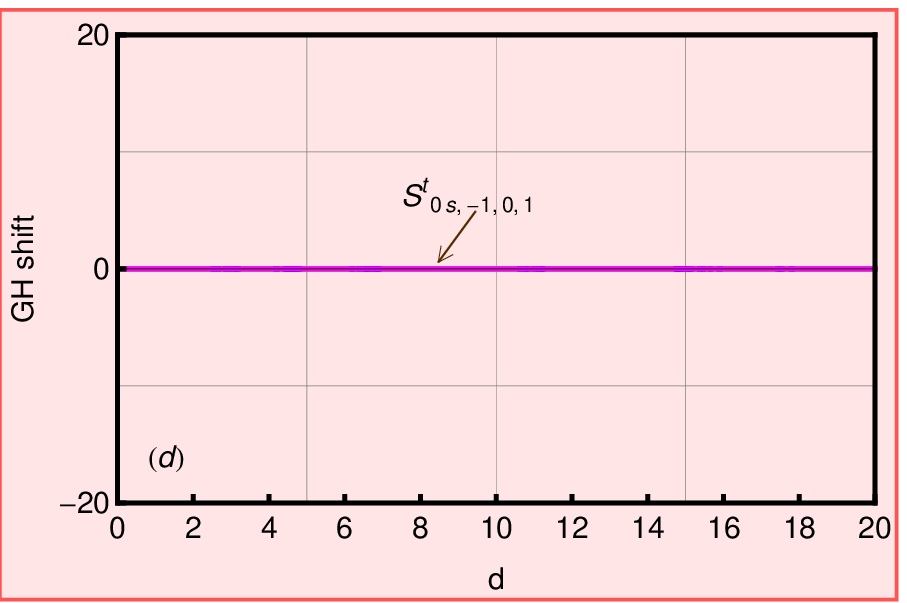}
 \caption{\sf{(color online)
 Transmission
 probability $T_{l}$ and GH shifts in transmission $S_{l}^{t}$ versus barrier width $d$, with $\alpha_1=0$ for static
 barrier, $\alpha_1=0.6$ for the oscillating barrier, $\epsilon=10$, $v=15$, $\mu=0$,$\varpi=1$. (a,c): $k_{y}=2$  and
(b,d): $k_{y}=0$. {For static barrier ($T_{0s}$, $S_{0s}^{t}$) in  magenta and for oscillating barrier ($T_{0}$, $S_{0}^{t}$) in blue,
($T_{-1}$, $S_{-1}^{t}$) in green, ($T_{1}$, $S_{1}^{t}$) in red}.}}\lb{figm2}
\end{figure}

\begin{figure}[!ht] \centering
\includegraphics[width=6cm, height=3.3cm]{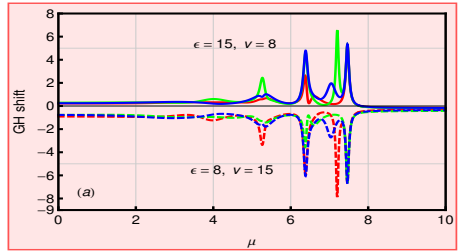}
\caption{\sf{(color online)  GH shifts in transmission
$S_{-1}^{t}$ (green line), $S_{0}^{t}$ (blue line), $S_{-1}^{t}$
(red line) versus energy gap $\mu$ with $\alpha_1=0.5$, $k_{y}=2$,
$d=1.5$, $\varpi=1$, $(\epsilon=8, v=15)$ and
$(\epsilon=15,
v=8)$.}}\lb{figm3}
\end{figure}

At this stage let us see what will happen if a gap is introduced in
the intermediate region ${\sf 1}$ ($0\leq x\leq d$). As is shown in Figure \ref{figm3} 
the gap affects
the system {energy spectrum}
obtained in region ${\sf 1}$. In fact, 
the GH
shifts $S_{l}^{t}$ in the propagating case can be enhanced by a
gap opening at the Dirac point. This has been performed by fixing
the parameters $\alpha_1=0.5$, $k_{y}=2$, $d=1.5$, $\varpi=1$ and
making different choices for the energy and potential. For the
configuration $(\epsilon=15, v=8)$, we can still have positive
shifts while for configuration $(\epsilon=15, v=8)$ the GH
shifts are negative. It is clearly seen that there are three intervals
showing different behaviors of the shifts. Indeed, for $\mu \in [0,4]$
the shifts are zero or  constants according to the energy configurations, but for $\mu \in [4,8]$ their behaviors changed completely
by exhibiting some peaks and vanish in the last interval $\mu\in[8,10]$. 

\begin{figure}[!ht] \centering
\includegraphics[width=6cm, height=3.3cm]{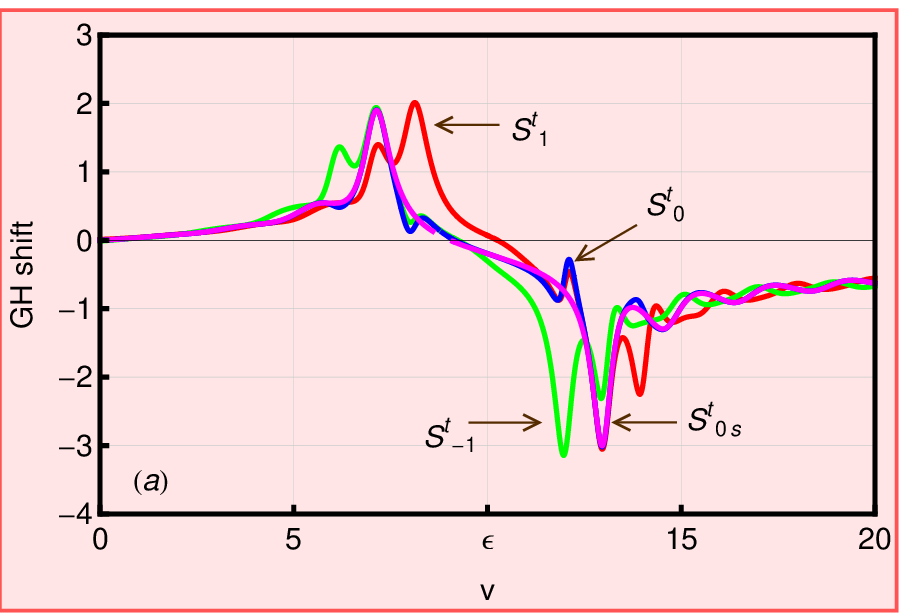}\ \ \ \
\includegraphics[width=6cm, height=3.3cm]{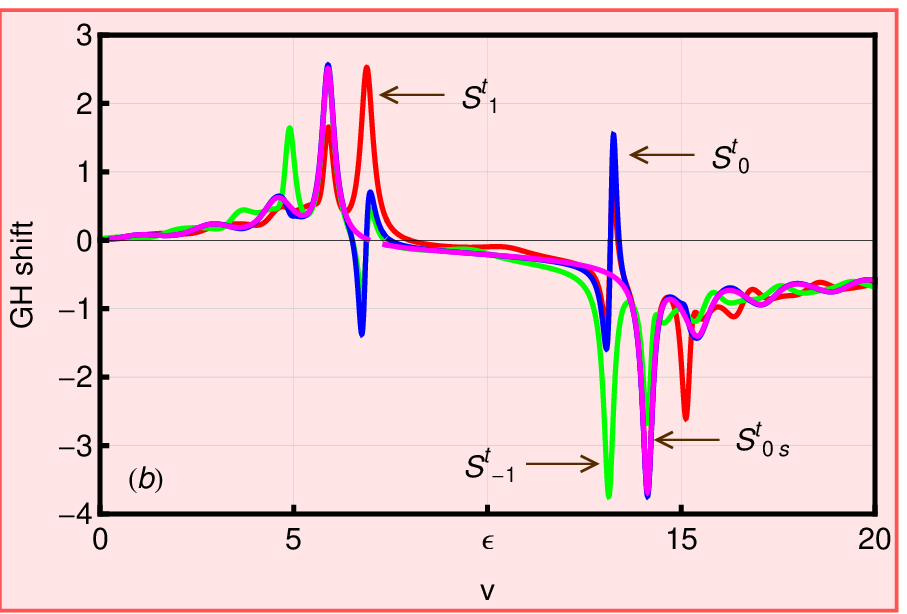}\\
\includegraphics[width=6cm, height=3.3cm]{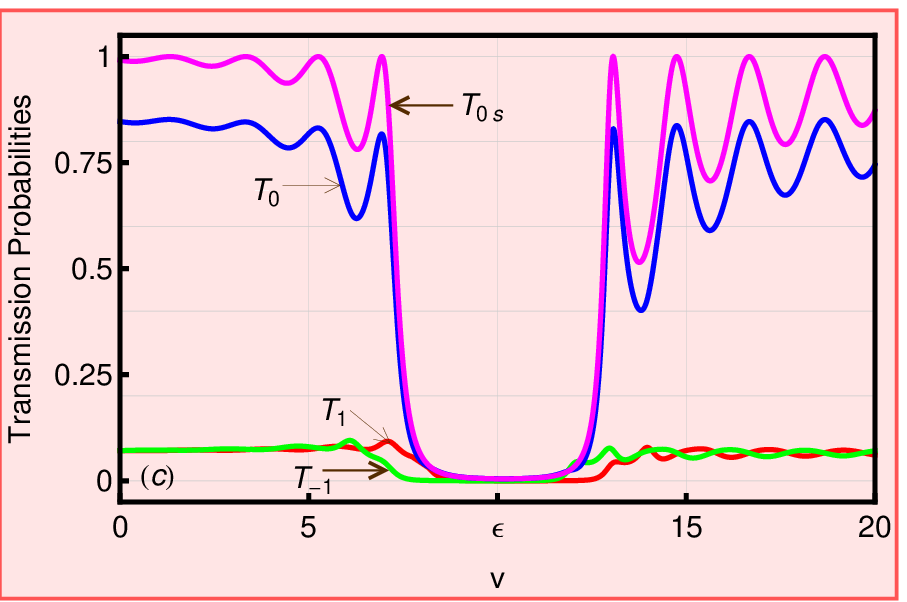}\ \ \ \
\includegraphics[width=6cm, height=3.3cm]{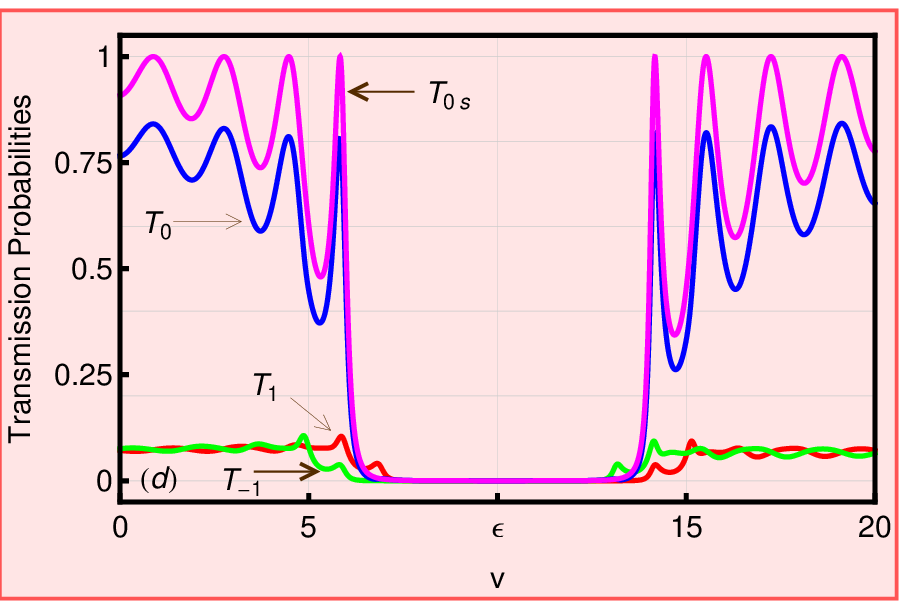}
 \caption{\sf{(color online) GH shifts in transmission and transmission
 probabilities for central band and first tow sidebands for $\alpha_1=0.4$ (oscillating barrier) along
with that for static barrier $\alpha_1=0$ as  function of the
potential $v$ with $\alpha_1=(0, 0.4)$
 $k_{y}=2$, $\epsilon=10$, $d=1.5$ ,$\varpi=1$. (a,c): $\mu=1$
 and (b,d):  $\mu=3$.
 {For static barrier ($T_{0s}$, $S_{0s}^{t}$) in  magenta and for oscillating barrier ($T_{0}$, $S_{0}^{t}$) in blue,
($T_{-1}$, $S_{-1}^{t}$) in green, ($T_{1}$, $S_{1}^{t}$) in red}.}}\lb{figm4}
\end{figure}

  Figure \ref{figm4} shows the GH shifts in transmission and
transmission probabilities for the central band and first two
sidebands for $\alpha_1=0.4$ (oscillating barrier) along with that
for static barrier $\alpha_1=0$ as  a function of the
potential $v$ for specific values 
$k_{y}=2$, $\epsilon=10$, $\varpi=1$
and different values of the energy gaps, see Figures \ref{figm4}(a,c) for $\mu=1$ and Figures \ref{figm4}(b,d)
for $\mu=3$. We observe that the region of 
weak GH shifts becomes wide with  increase in energy gap $\mu$,
the shifts are also affected by the parameters of the single
barrier. In particular it changes the sign at the total reflection
energies and peaks at each bound state associated with the
barrier. Thus the GH shifts can be enhanced by the presence of
resonant energies in the system when the incident angle is less
than the critical angle associated with total reflection. It is
clearly seen that  $ S_{l}^{t}$ are oscillating
between negative and positive values around the critical point
$v=\epsilon+l\varpi$ $(l=0,\pm 1)$. At such point $T_{l}$ is
showing transmission probabilities for the central band and first two
sidebands while it oscillates away from the critical point. We notice that for
large values of $v$, the GH shifts become mostly constant
and
can be positive as well as negative. We deduce that there
is a strong dependence of the GH shifts on the potential height
$v$, which can help to realize a controllable 
{sign of the} GH shifts.

\begin{figure}[!ht] \centering
\includegraphics[width=6cm, height=3.3cm]{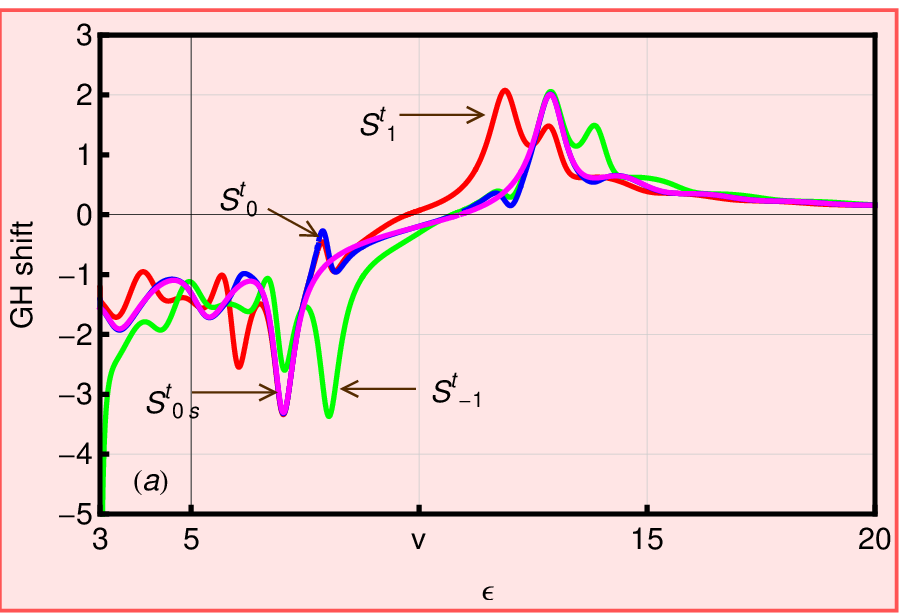}\ \ \ \
\includegraphics[width=6cm, height=3.3cm]{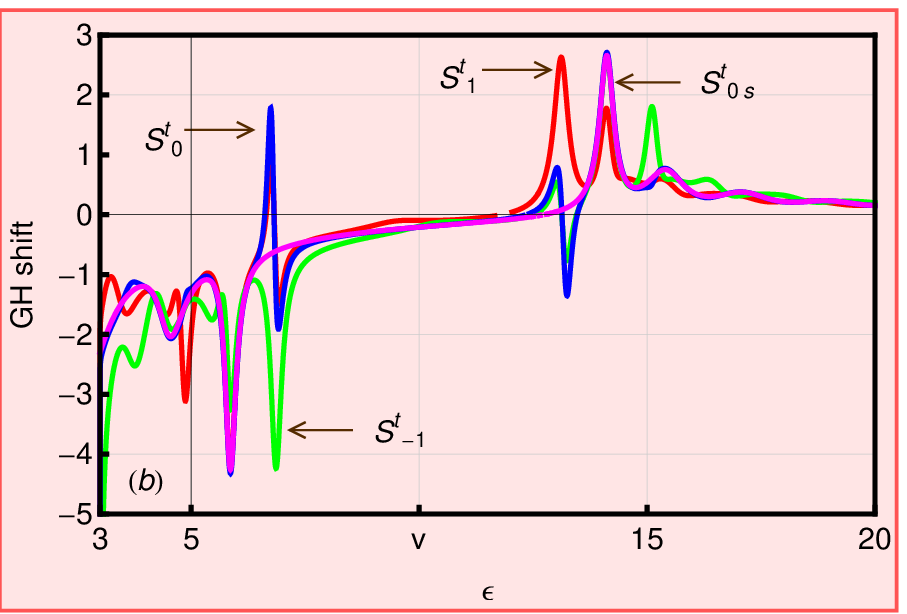}\\
\includegraphics[width=6cm, height=3.3cm]{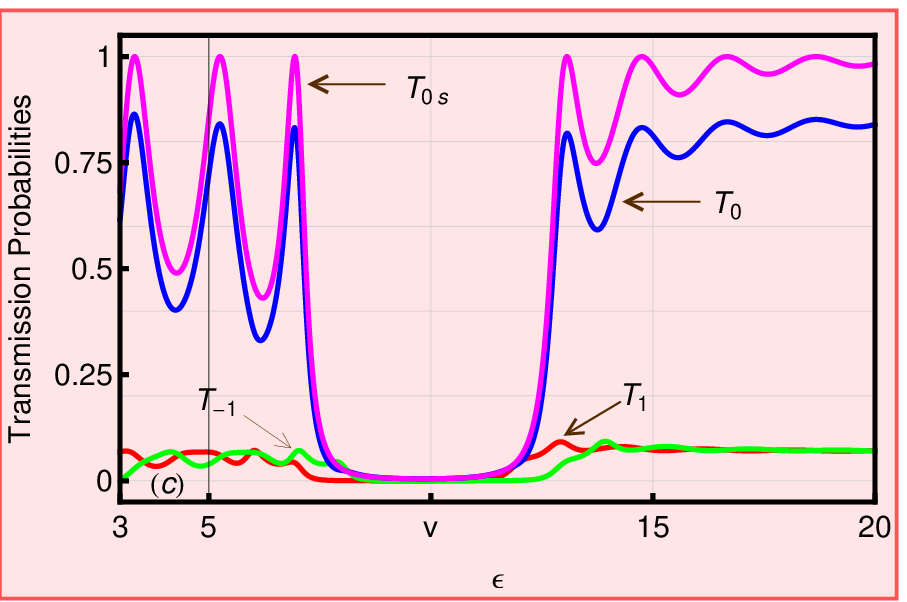}\ \ \ \
\includegraphics[width=6cm, height=3.3cm]{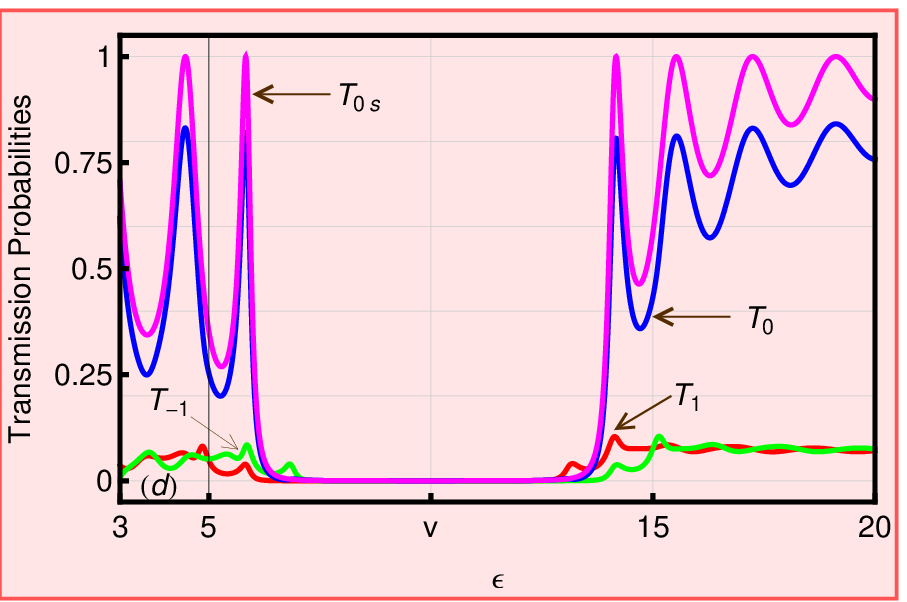}
 \caption{\sf{(color online) GH shifts  in transmission $S_{l}^{t}$ and transmission
 probabilities $T_l$ as function of the incident energy $\epsilon$ with $\alpha_1=(0, 0.4)$,
 $k_{y}=2$, $v=10$, $d=1.5$, $\varpi=1$. (a,c): $\mu=1$  and (b,d): $\mu=3$.
 {For static barrier ($T_{0s}$, $S_{0s}^{t}$) in  magenta and for oscillating barrier ($T_{0}$, $S_{0}^{t}$) in blue,
($T_{-1}$, $S_{-1}^{t}$) in green, ($T_{1}$, $S_{1}^{t}$) in red}.}}\lb{figm5}
\end{figure}

From Figure \ref{figm5}, we see that the GH shifts in transmission
$S_{l}^{t}$ and the transmission probabilities $T_l$ versus
incident energy $\epsilon$ for the values $\alpha_1=(0, 0.4)$
 $k_{y}=2$, $v=10$, $d=1.5$, $\varpi=1$ for two values of gap  $\mu=1$ 
 in Figure \ref{figm5}(a,b)  and
 $\mu=3$ Figure in \ref{figm5}(b,d). Both quantities are showing a series of peaks
and resonances where the resonances correspond to the bound states of
the static
 barrier for $\alpha_1=0$ and the oscillating barrier for $\alpha_1=0.4$. We notice that the GH shifts in transmission peak at each
bound state energy are clearly shown in the transmission curve
underneath. The energies at which transmission vanishes correspond
to energies at which the GH shifts in transmission change sign.
Since these resonances are very sharp (true bound states with zero
width) it is numerically very difficult to track all of them, if
we do then the alternation in sign of the GH shifts will be
observed. We observe that  around the Dirac point
$\epsilon=v+l\varpi$ the number of peaks is equal to the number of
transmission resonances.

\section{Conclusion}

We have studied the Goos-H\"anchen shifts for
Dirac fermions in gapped graphene through single barrier with a time periodic
potential.
This has been done  using the solutions of the energy spectrum
to write down {the incident, reflected and transmitted beams}
in integral forms. In the second step, we have employed our results
to determine the phase shifts associated {with} the transmission and reflection amplitudes. Subsequently, we have  derived 
the corresponding GH shifts in terms of various physical parameters
such as the width and height of the barriers, incident energy, transverse wavevector and frequency of {oscillating barrier}.

The time periodic electrostatic potential generates additional
sidebands at energies $\varepsilon+l\omega$ $(l=0,\pm 1, \pm 2, \cdots)$
in the transmission probability originating from photons absorption or emission within the oscillating barrier. However, at low energies we can limit
ourselves to single photon processes and neglect two photon processes.
Numerically, we have shown that the GH
shifts for the central band and first two sidebands depend on the incident
angle of the particles, the width and height of the barrier, and
the frequency of oscillation.
Our results showed that the GH
shifts are affected by the internal structure of the 
oscillating barrier. We have analyzed the GH shifts in the 
transmission in terms of incident energy, barrier width, potential
strength
and energy gap. We have observed that the GH shifts in the transmission for the 
central band and first two sidebands change sign at the Dirac points
$\epsilon=v+l\omega$ and switch from positive to negative
signs in a controllable manner.
The energies at which the GH shifts in transmission change sign correspond to the sharp resonances at which the transmission vanishes. Thus, the switching of the sign of the GH shifts can be selected  in a controllable manner. Then our results might provide a theoretical basis for the design of graphene based electronic switches or high sensitivity sensors based on the sign of the GH shifts.

Finally, we mention that our results could be experimentally tested 
using
a beam splitter scanning method \cite{Xin}, 
which allows to 
 measure the giant GH shifts produced by increasing the thicknesses of gapped graphene
subject to a time oscillating potential.

\section*{Acknowledgments}

The generous support provided by the Saudi Center for Theoretical
Physics (SCTP) is highly appreciated by all authors. AJ and HB
acknowledge the support of KFUPM under research group project RG181001.


\end{document}